\documentclass[review]{elsarticle}
\usepackage{graphicx}
\usepackage[figurename=Fig.,bf]{caption}
\usepackage{epsfig}
\usepackage{subfig}
\usepackage{amsmath}

\begin{document}
\begin{frontmatter}

\address[famu]{Florida A\&M University,Department of Physics,
Tallahassee FL,32307}

\title{Toroidal moments of Schr\"{o}dinger eigenstates}

\author [famu]{M. Encinosa\corref{cor1}}
\ead{mario.encinosa@famu.edu}
\author [famu]{J. {Williamson}}
\ead{johnny.williamson@cepast.famu.edu}

\cortext[cor1]{Corresponding Author}

\begin{abstract}
The Hamiltonian for a particle constrained to motion near a toroidal helix
with loops of arbitrary eccentricity is developed. The resulting
three dimensional Schr\"{o}dinger equation is reduced to a one dimensional
effective equation inclusive of curvature effects. A basis set
is employed to find low-lying eigenfunctions of the helix.
Toroidal moments corresponding to the individual eigenfunctions are calculated.
The dependence of the toroidal moments on the eccentricity of the
loops is reported. Unlike the classical case, the moments strongly depend
on the details of loop eccentricity.
\end{abstract}


\end{frontmatter}

\bibliographystyle{elsarticle-num}

\section{Introduction}
The majority of work directed towards modeling the metaparticle
constituents of metamaterials has been performed using classical
physics \cite{wegener}. The characteristic length scales of most currently
fabricated metaparticles allow for that approach to be appropriate
and productive. However, it is nearly certain that metaparticles will
eventually be fabricated on scales at which quantum mechanical
methods will prove necessary to capture their physics with good
fidelity \cite{zhang, shea, lorke}.

This paper focuses upon two interesting properties common to many
metaparticles: they can be approximated as reduced dimensionality
systems and they can possess nontrivial topologies. The advent of
quasi one and two dimensional curved nanostructures has led to
situations wherein formalism  developed for particles
constrained to curved manifolds has become of practical importance. 
Specifically, there exits
a prescription that allows for degrees of freedom
 extraneous to the particle's 'motion' on a curve or
surface to be shuttled into effective curvature potentials in the
Schr\"{o}dinger equation \cite{chapblick, dacosta1,dacosta2,duclosexner,ee1,ee2,jenskoppe,matsutani1,matsutani2,taira}.

Recently, it was suggested that quantum methods be employed in an
effort towards understanding toroidal moments induced by currents
supported on nanoscale metaparticles and the interactions of those
moments with time-dependent electromagnetic fields \cite{kaelscience}. Because
of the theoretical and practical interest in toroidal moments \cite{afanasiev, ceulemans,
dubovik, papas, naumov,spaldin, sawada},
 a toroidal helix (TH) of adjustable eccentricity has been
chosen here to investigate the role of quantum effects.
Being closed, a TH can support current carrying solutions allowing
for the existence of a toroidal moment \cite{kibis}. Furthermore, the TH has
the advantage of having sufficient symmetry to allow for a clean
reduction of the full Hamiltonian to a one dimensional effective
Hamiltonian.

The goals of this work are threefold. The first is to derive the
Hamiltonian for a particle in a coordinate system adapted
to include points near the coils of a TH of arbitrary
eccentricity. The next deals with reducing the full three
dimensional Hamiltonian via a well known procedure \cite{dacosta1,schujaff,burgsjens} to arrive
at an effective one-dimensional Schr\"{o}dinger equation. The
reduction of dimensionality impels the introduction of a curvature
potential well known to workers in the field of curved
manifold quantum mechanics. A basis set consistent with the
periodicity and symmetry of the system is introduced thereafter.
Achieving the first two goals and with the basis functions in
hand, the spectrum and wave functions of the system (which can
 be used for applications in the external field and/or
time-dependent case) are found. Finally, toroidal moments
corresponding to particular eigenstates are determined and their
sensitivity to the eccentricity of the loops comprising the TH is
investigated.

The remainder of this paper is organized into four sections.
Section 2 introduces a parameterization for an $\omega$ turn TH
in terms of an azimuthal coordinate $\phi$. A three
dimensional Hamiltonian $H^3_{\omega}$ appropriate to motion near
the TH follows by attaching a Frenet system to the helix and
assigning two coordinates $q_N,q_B$ to describe degrees of freedom
away from the coil. Section 3 details the reduction of
$H^3_{\omega}$ to a one-dimensional $H^1_{\omega}$ by standard
methods, although perhaps unfamiliar to workers in the metamaterial
community. As a consequence of the reduction, curvature potentials
appear. Their presence has been shown to be essential in properly
describing one dimensional systems that exist in an ambient higher
dimensional space \cite{enchardsoft}. Section 4 presents the basis set used to
calculate the spectrum, eigenstates and toroidal moments per a
given quantum state. Those quantities, along with results showing
the dependence of TMs on eccentricity are given. Section 5 is
dedicated to conclusions and some remarks concerning future work.

\section{The TH Schr\"{o}dinger equation}

To arrive at the time independent Schr\"{o}dinger equation
$H^3_{\omega}({\bf
  r})\Psi= E\Psi = \big(-{1\over 2}\nabla^2 +V \big )\Psi $, the Laplacian must be derived
  from a suitable parameterization of the TH geometry.
Consider a TH with $\omega$ equally spaced circular coils.  Let
\textit{R} be the distance from the z-axis to a loop center and
\textit{a} the radius of a loop. First define
\begin{equation}
W(\phi)=R+a~{\rm cos}(\omega\phi)
\end{equation}
with $\phi$ the usual cylindrical coordinate azimuthal angle. The
 circular TH  is traced out by the Monge form \cite{graustein}
\begin{equation}
{{\bf{r}}}(\phi)=W(\phi)\hat{{\boldsymbol {{\boldsymbol {\rho}}}}}+a~{\rm sin}(\omega\phi)\hat{{\bf{k}}}.
\end{equation}
Generalizing Eq.(2) to coils of arbitrary eccentricity requires
only the modification
\begin{equation}
{\bf{r}}(\phi)=W(\phi)\hat{{\boldsymbol {\rho}}}+b~{\rm sin}(\omega\phi)\hat{{\bf{k}}}
\end{equation}
where  $a,b$ may be adjusted to yield the coil shape desired
(Fig. 1). To avoid cluttering the narrative with blocks of
equations, the expressions that follow will apply to the circular
case only. The expressions for  arbitrary $a$ and $b$ are given in the
appendix.

A three dimensional neighborhood in the vicinity of the TH is
built by assigning two coordinates to points near the curve along
unit vectors orthogonal to the curve's tangent and to each other. The
Frenet-Serret equations \cite{graustein} provide such an orthonormal coordinate
system known as a Frenet trihedron. The unit tangent to any
point on a curve traced  by ${\bf{r}}(\phi)$ is
\begin{equation}
\hat{{\bf{T}}}={d{\bf{r}}(\phi)\over{d\phi}}~{\bigg| \bigg| {d{\bf{r}}(\phi)\over{d\phi}} \bigg| \bigg|}^{-1}
\end{equation}
\noindent from which the Frenet trihedron can be constructed via the relations
\begin{equation}
{d\hat{{\bf{T}}}\over {d\phi}}= {\bigg| \bigg| {d{\bf{r}}(\phi)\over{d\phi}} \bigg| \bigg|} \kappa(\phi) \hat{{\bf{N}}}
\end{equation}
\begin{equation}
{d\hat {{\bf{N}}}\over {d\phi}}= {\bigg| \bigg| {d{\bf{r}}(\phi)\over{d\phi}} \bigg| \bigg|} \big(  -\kappa(\phi) \hat{{\bf{T}}}+\tau(\phi)
{\hat{{\bf{B}}}} \big)
\end{equation}
\begin{equation}
{d\hat{{\bf{B}}}\over {d\phi}}=- {\bigg| \bigg| {d{\bf{r}}(\phi)\over{d\phi}} \bigg| \bigg|} \tau(\phi) \hat{{\bf{N}}}
\end{equation}
where the curvature and torsion of the space curve ${\bf{r}}(\phi)$
are indicated by $\kappa(\phi)$ and $\tau(\phi)$ respectively (where again, detailed forms for
the expressions in Eqs. (4-7) appear in the appendix).  Points near the TH are located
via two perpendicular displacements $q_N\hat{{\bf{N}}}$ and $q_B\hat{{\bf{B}}}$.
The TH position vector may now be written
\begin{equation}
{\bf{x}}(\phi,q_N,q_B)={\bf{r}}(\phi)+q_N\hat{{\bf{N}}}+q_B\hat{{\bf{B}}}.
\end{equation}
\noindent It should be noted that Eq.(8) defines a Cartesian
region about a curve traced by ${\bf{r}}(\phi)$. While it is
certainly possible to construct a finite tubular neighborhood
about  ${\bf{r}}(\phi)$,  the coordinate ambiguity of the azimuthal
angle as the radial distance approaches zero causes the limiting
procedure to become complicated. Additionally, the separability of the
Schr\"{o}dinger equation into tangential and normal variables is lost,
and with it any real advantage in using the reduced Hamiltonian.

The covariant metric tensor elements $g_{ij}$ can be read off of
the quadratic form \cite{arfkenweber}
\begin{equation}
d{\bf{x}} \cdot d{\bf{x}}=g_{ij}dq^idq^j
\end{equation}
where in what follows the ordering convention is $(q^1,q^2,q^3)
\equiv (\phi,q_N,q_B)$. The Laplacian is
\begin{equation}
\nabla^2={1 \over \sqrt{g}}{\partial \over{\partial q^i}}
\bigg(\sqrt{g}~g^{ij} {\partial \over {\partial q^j}}\bigg)
\end{equation}
with $g=det(g_{ij})$ and $g^{ij}$ the contravariant components of
the metric tensor. Before presenting explicit forms for
$g_{ij}$ and $g^{ij}$ , it is useful to define
\begin{equation}
f(\phi)=\bigg | \bigg | {{d \bf {r}(\phi)} \over{{d\phi}}} \bigg | \bigg | =[(a
\omega)^2 + W(\phi)^2]^{1/2}
\end{equation}
and
\begin{equation}
G(\phi,q_N)=1-q_N \kappa(\phi)
\end{equation}
after which the covariant metric may be written \vskip 2 pt
\begin{equation}
g_{ij}=\begin{pmatrix} f(\phi)^2[G(\phi,q_N)^2+\tau(\phi)^2(q_N^2+q_B^2)]
& -\tau(\phi) q_B f(\phi)& \tau(\phi) q_N f(\phi) \\ -\tau(\phi) q_B f(\phi) & 1 & 0 \\
\tau(\phi) q_N f(\phi) & 0 & 1\end{pmatrix}.
\end{equation}
\vskip 6pt \noindent The contravariant form of the metric is
obtained straightforwardly; \vskip 2pt
\begin{equation}
g^{ij}={1\over{f(\phi)^2G(\phi,q_N)^2}}\begin{pmatrix}{1} & {\tau(\phi)
q_B f(\phi)} & {-\tau(\phi) q_N f(\phi)} \\ {\tau(\phi) q_B f(\phi)} & f(\phi)^2[G(\phi,q_N)^2
+{\tau(\phi)^2 q_B^2]} & -{\tau(\phi)^2 q_N q_B f(\phi)^2} \\ {-\tau(\phi)
q_N f(\phi)} & -{\tau(\phi)^2 q_N q_B f(\phi)^2} &
f(\phi)^2[G(\phi,q_N)^2+{\tau^2(\phi) q_N^2]}\end{pmatrix}.
\end{equation}
It is easy to show that
$$\sqrt{g}=f(\phi)\big( 1- q_N\kappa(\phi) \big).$$
\vskip4pt \noindent The Laplacian found by directly evaluating
Eq.(10) is complicated by the existence of cross terms arising from ${\partial^2 /{
\partial q^i
\partial q^j}}$, ($i\neq j$), operations. However, all of those terms are multiplied
 by the distance parameters $q_N$ and $q_B$  such that in the limit $q_N,
q_B \rightarrow0$ they vanish independently of the derivative operators that
follow them. Taking this limit now (it will be taken again later
post operation of the $q_{N,B}$ derivatives) leads to a more
convenient starting point for developing the reduced Hamiltonian
in the ensuing section. Physically, the limiting procedure is
effected by external mechanical or electrical constraints; mathematically,  they are added
by hand into the Schr\"{o}dinger equation as potentials $V_n(q)$
normal to the lower dimensionality base manifold. Their detailed forms are not important.
Previous work has shown that even for finite thicknesses, degrees
of freedom  extraneous to those of the base manifold do not mix
with the latter in the sense that their wave functions decouple
\cite{enchardsoft}. Here, for the sake of definiteness, hard wall potentials are
assumed for $V_n(q_{\mbox{\tiny N}})$ and $V_n(q_{\mbox{\tiny B}})$ in this and the next section. With
this discussion in mind, $H^3_\omega$ may be written as (with $\hbar=m=1$)
\begin{equation}
H^3_\omega=-{1\over{2}}  \bigg({1 \over{f(\phi)^2}}{\partial^2
\over{\partial \phi^2}}-{f'(\phi) \over{f(\phi)^3}}{\partial
\over{\partial \phi}}- \kappa(\phi){\partial
\over{\partial q_N}}+{\partial^2 \over{\partial
q_N^2}}+{\partial^2 \over{\partial q_B^2}}~\bigg)+
 V_n(q_N)+V_n(q_B).
\end{equation}

\noindent Note that the $H^3_\omega$ at this stage is still not
separable. The procedure for rendering $H^3_\omega$ separable and
arriving at a simpler effective Hamiltonian is given in the
following section.

\section{Constructing  the effective Hamiltonian}
As the particle is constrained to the toroidal helix, its wave function will decouple into tangent and normal functions
(the subscripts \textit{t} and \textit{n} denote tangent and normal respectively)
\begin{equation}
\Psi (\phi,q_N,q_B) \rightarrow \chi_t (\phi) \chi_n (q_N) \chi_n (q_B)
\end{equation}
and $G(\phi,q_N)$ will approach unity.
The normalization condition
\begin{equation}
\int_0^{2 \pi} \vert\Psi(\phi,q_N,q_B)\vert^2 G(\phi,q_N)f(\phi)\,d\phi\,dq_N\,dq_B=1
\end{equation}
becomes
\begin{equation}
\int_0^{2 \pi} \vert \chi_t (\phi) \vert^2~\vert \chi_n (q_N) \vert^2~\vert \chi_n (q_B) \vert^2 f(\phi)\,d\phi\,dq_N\,dq_B=1.
\end{equation}
The norm must be conserved in the decoupled limit \cite{dacosta1}, which implies
\begin{equation}
\vert\Psi(\phi,q_N,q_B)\vert^2 G(\phi,q_N)=\vert \chi_t (\phi) \vert^2~\vert \chi_n (q_N) \vert^2~\vert \chi_n (q_B) \vert^2.
\end{equation}
The wave function $\Psi(\phi,q_N,q_B)$ is now related to $\chi_t (\phi) \chi_n (q_N) \chi_n (q_B)$ by
\begin{equation}
\Psi(\phi,q_N,q_B)=\chi_t (\phi) \chi_n (q_N) \chi_n (q_B) G^{-{1 /2}}(\phi,q_N).
\end{equation}
Applying $H^3_\omega$  to $\Psi(\phi,q_N,q_B)$ and  taking the limit as $q_N, q_B \rightarrow0$ post all derivative operations yields the result
\begin{equation}
H^3_\omega=-{1 \over{2}}\bigg({1 \over{f(\phi)^2}}{\partial^2 \over{\partial \phi^2}}-{f'(\phi) \over{f(\phi)^3}}{\partial \over{\partial \phi}}+{1\over{4}}\kappa^2(\phi) +{\partial^2 \over{\partial q_N^2}}+{\partial^2  \over{\partial q_B^2}}~\bigg)+V_n(q_N) +V_n(q_B).
\end{equation}
Distributing the energy  between the  $(\phi,q_N,q_B)$ degrees of freedom by allowing $E=E_{\phi}+E_N+E_B$, leads to the decoupled system
\begin{equation}
-{1 \over{2}}\bigg({1 \over{f(\phi)^2}} {\partial^2 \over{\partial \phi^2}}-{f'(\phi) \over{f(\phi)^3}} {\partial \over{\partial \phi}}+{1\over{4}} \kappa^2(\phi) \bigg) \chi_t(\phi)=E_\phi \chi_t(\phi)
\end{equation}
\begin{equation}
-{1 \over{2}}~{\partial^2 \chi_n(q_N) \over{\partial q_N^2}} + V_n(q_N) \chi_n(q_N)=E_N \chi_n(q_N)
\end{equation}
\begin{equation}
-{1 \over{2}}~{\partial^2 \chi_n(q_B) \over{\partial q_B^2}} + V_n(q_B) \chi_n(q_B)=E_B \chi_n(q_B).
\end{equation}

\noindent Since $V(q_N)$ and $V(q_B)$ are the confining potentials effecting the $q_N, q_B \rightarrow0$ constraint, $q_N$ and $q_B$ can be considered spectator variables and only the $\phi$-dependent part of the Hamiltonian indicated in Eq.(21) is nontrivial. The Hamiltonian in one dimension $H^1_\omega$ is written

\begin{equation}
H^1_{\omega}=-{1 \over{2}}\bigg({1 \over{f(\phi)^2}} {\partial^2 \over{\partial \phi^2}}-{f'(\phi) \over{f(\phi)^3}} {\partial \over{\partial \phi}} \bigg)+V_c(\phi)
\end{equation}
\noindent
with
\begin{equation}
V_c (\phi)=-{1\over{8}} \kappa^2 (\phi)
\end{equation}
\noindent the curvature potential. The curvature potential $V_c(\phi)$ emerges as an artifact of embedding the particle's one dimensional path of motion in the ambient three dimensional space. The explicit form of the curvature potential in Eq.(26) can be determined from
\begin{equation}
\kappa (\phi)=[P_1(\phi)^2+P_2(\phi)^2]^{1/2}
\end{equation}
where
\begin{equation}
P_1(\phi)=-{{a\omega^2 + W(\phi){\rm cos}(\omega\phi)}\over{f(\phi)^2}}
\end{equation}
and
\begin{equation}
P_2(\phi)= \frac{{\rm sin}(\omega\phi)}{f(\phi)}\bigg[1+\bigg({a\omega \over{f(\phi)}}\bigg)^2~\bigg].
\end{equation}
\noindent Explicit forms of the tangent, normal, and binormal vectors, along with other vectors and functions for the circular and elliptic helices are given in the appendix.

A plot of $V_c(\phi)$ for some representative values of $a,b$ with $\omega = 4$ appears in Fig. 2. Note that the circular case values are negligible in magnitude compared to the eccentric cases, and when $a > b$, $V_c(\phi)$ is substantially larger than for the converse. For  larger ratios of $a$ to $b$, $V_c(\phi)$ can be orders of  magnitude larger than indicated in the figure.

 It is worth stating that instead of parameterizing the TH with $\phi$, it would also be possible to employ an arc length scheme where an arc length parameter $\lambda$ is determined from $\lambda=\int_0^{\phi}f(\phi') \,d\phi'.$ However, to include the curvature potential as a function of $\lambda$, it would be necessary to find $\phi(\lambda)$ along the curve. While this could be accomplished numerically, using the azimuthal angle is somewhat better suited to incorporating external fields \cite{encarbB,encdipole}.

\section{Computational methods and results }
If the TH is small enough to require a quantum mechanical
description, the $\phi$-dependent part of its wave function must obey Bloch's theorem (the $t$-subscript will be dropped hereafter)
\begin{equation}
\chi_k\bigg(\phi+{2\pi \over \omega}\bigg) ={\rm exp}\bigg[ik{2\pi
\over \omega}\bigg]\chi_k(\phi).
\end{equation}
A standard choice is \cite{grossoparra}
\begin{equation}
\chi_k(\phi) ={\rm exp}(ik\phi) u_k(\phi)
\end{equation}
\noindent where $u_k(\phi+2\pi /  \omega)=u_k(\phi)$ is satisfied. Single
valuedness requires the Bloch index $k \equiv p$ = integer. A
convenient choice for $u_p(\phi)$ basis elements is
\begin{equation}
u_p(\beta,\phi)={\rm exp}[i\beta\phi].
\end{equation}
The requirement indicated in Eq.(30) yields
\begin{equation}
\beta=\omega n, \ \ \ \   n \equiv {\rm integer}.
\end{equation}
From the above considerations, a suitable basis set for the TH is
\begin{equation}
\chi^{p\alpha}(\phi)={\rm exp}[i p \phi]\sum_n C^{p\alpha}_n {\rm exp} (i n \omega
\phi).
\end{equation}
\noindent
The Bloch form introduces sub-states (sub-bands in the
case of a continuous rather than discrete index) for each $p$ value which would
not be present if the TH were treated as a ring of length $L$. The $C^{p\alpha}_m$ are the expansion coefficients for $\alpha$-th sub-state of a given $p$ value.
In this work, it was found that a five-state expansion proved sufficient
to yield basis size independent results for the lower $p$ sub-states.
For $\omega$ turns, values of $p$  consistent with the Bloch
theorem, $ p < \omega$, are used. For clarity, only $p \geq 0$
are discussed.

A disadvantage of directly adopting the expression given by Eq.(34) is that the basis functions are not orthogonal over the
integration measure $f(\phi) d\phi$. A more
natural basis set is given by a re-scaled form of Eq.(34)
\begin{equation}
\chi^{p\alpha}(\phi)={{\rm exp}[i p \phi]\over {f(\phi)^{1 / 2}}}\sum_n
C^{p\alpha}_n {\rm exp} (i n \omega \phi).
\end{equation}
With basis function orthogonality preserved on the right hand side
of the Hamiltonian, eigenvalues and eigenvectors are calculated by
diagonalizing the matrix comprising the elements
\begin{align}
\begin{split}
H_{mn}={1 \over 2\pi}\int_0^{2\pi} &e^{i\omega(n-m)\phi}\big[-2 V_c
(\phi) -\frac{(p+\omega n)^2}{f(\phi)^2} +\frac{5}{4} \frac{f'(\phi)^2}{f(\phi)^4} -\frac{f''(\phi)}{2 f(\phi)^3} \\ &-2 i (p+ \omega n) \frac{f'(\phi)}{f(\phi)^3} \big] \,d\phi.
\end{split}
\end{align}
Once the eigenstates are found, the current in general is calculated with
(now with units)
\begin{equation}
{\textbf{\textit{j}}}(\phi,q_B,q_N)={q_e \hbar \over{m_e}}Im \big [\Psi^\ast(\phi,q_B,q_N) {\boldsymbol{\nabla}}\Psi(\phi,q_B,q_N)
\big ].
\end{equation}
The current density given by Eq.(37) is inclusive of cross-sectional degrees of freedom and yields a current passing through a rectangular area with unit
 normal $\hat{{\bf{T}}}$. However, in keeping with the intent of this work, the limit of infinitesimal thickness is assumed (or equivalently, the $q_B, q_N$ degrees of freedom are integrated out) leading to the current expression for the $p\alpha$-th state
\begin{equation}
{\textbf{\textit{j}}}^{p\alpha}(\phi,0,0)={{q_e \hbar } \over {m_e}}Im \bigg [{(\chi^{p\alpha}(\phi))^\ast \over f(\phi)}  { \partial \chi^{p\alpha}(\phi) \over \partial
\phi }\bigg ]{\hat{\bf{T}}}
\end{equation}
\noindent where the form of the reduced gradient operator is obvious. The  quantum mechanical current that stems from Eqs.(35) and (38) becomes
\begin{equation}
{\textbf{\textit{j}}}^{p\alpha}(\phi,0,0)={q_e \hbar \over{m_e}}{1 \over{2\pi}}\sum\limits_{m,n} C^{p\alpha}_m C^{p\alpha}_n
\bigg[{(p+\omega n) \over{f(\phi)^2}} {\rm cos}[\omega(n-m)\phi]-{f'(\phi)\over{2f(\phi)^3}} {\rm sin}[\omega(n-m)\phi]\bigg]~\hat{{\bf{T}}}.
\end{equation}
When $V_c(\phi)$ is included in the Hamiltonian  the $C^{p\alpha}_m$ are modified, causing the current to become inclusive of curvature effects. This current is then used to calculate the toroidal moments according to \cite{marinov}
\begin{equation}
\textbf{T}^{p\alpha}_M={1 \over{10}}\int_0^{2\pi} [({\textbf{\textit{j}}}^{p\alpha}(\phi,0,0) \cdot {\bf{r}}) {\bf{r}} -2r^2 {\textbf{\textit{j}}}^{p\alpha}(\phi,0,0)~]\,f(\phi)d\phi.
\end{equation}
\noindent
Equation (40) allows calculation of quantum mechanical toroidal moments of ground and excited states for each Block index $p$. For a macroscopic thin wire where ${\textbf{\textit{j}}}d\tau \rightarrow Id{\bf{r}}$ is applicable, the toroidal moment for each $p$ reduces to the classical result
\begin{equation}
\textbf{T}^p_M={I \over{10}}\int_0^{2\pi} \bigg[ ({d{\bf{r}} \over{d\phi}} \cdot {\bf{r}}~){\bf{r}}-2r^2{d {\bf{r}} \over{d\phi}}~\bigg ] f(\phi)\,d\phi.
\end{equation}
For circular TH, Eq.(41) yields
\begin{equation}
\textbf{T}^p_M=-{{\pi \omega I a^2 R} \over{2}} ~\hat{{\bf{k}}}
\end{equation}
and for the elliptic TH,
\begin{equation}
\textbf{T}^p_M=-{{\pi \omega I a b R} \over{2}} ~\hat{{\bf{k}}}.
\end{equation}
As a means of comparison, the current for the $p$ state without curvature effects (i.e. a free particle on a given $\omega$ turn helix) is easily determined to be
\begin{equation}
I={2  \pi q_e \hbar p \over m_e L^2}
\end{equation}
where the total length of the TH, $L$, is calculated using $$L=\int_0^{2\pi} f(\phi) \,d\phi.$$

The formalism described in this section was employed to calculate
the eigenvalues and eigenstates expressed in terms of the $C^{p\alpha}_m$
for several $\omega$ and $p$ values. To get a sense of the modifications arising from
$V_c(\phi)$, the eigenvalues and amplitudes for a six-turn eccentric helix in a $p = 1$
state are listed without (Table 1) and with (Table 2) the curvature potential being
present. The eigenvalue shifts reflect that $V_c(\phi)$ is always attractive as shown in Fig. (2), and capable
of causing amplitude shifts. The reader will note there is no table
indicating the shifts for the circular case; \textit{the effects are negligible and essentially
independent of the coil radius} $a$.

 With the $C^{p\alpha}_m$ amplitudes in hand, Eq.(39) can be used to find the ${\textbf{\textit{j}}}^{p\alpha}(\phi,0,0)$
necessary for computing TMs. To set a baseline for understanding the effect of including
$V_c(\phi)$, the curvature potential was shut off and ${\textbf{\textit{j}}}^{p\alpha}(\phi,0,0)$
determined for many cases. In Fig. (3), representative results are given for $\omega = 4$, p = 1.
As anticipated, the lowest energy states yield very steady currents;  oscillations begin to
manifest in the higher sub-band energy states. Turning the potential on produces very little
change in the currents; in Fig. (4), it becomes clear that $V_c(\phi)$ does little,
 which is consistent with its small amplitude indicated in Fig. (2).

When eccentricity is introduced by setting $a=.75$ and $b=.25$, the results are less trivial.
The results displayed in Figs. (5) and (6) are representative of a general trend
observed throughout values of $\omega,p$. The curvature potential suppresses the current in every sub-band
by a discernable fraction. Similar behavior is observed when $a=.25$ and $b=.75$ as shown in Fig. (7)
and (8), but note that the magnitudes are substantially
different from the converse values of $a$ and $b$. The Bloch form of the wave function,
independent of the presence of $V_c(\phi)$ (which again lessens the magnitude of the currents), the Bloch form of the
wave function and the $ab$ dependence  of the Laplacian
are sufficient to cause asymmetries in the currents.

Toroidal moment results for $\omega = 4$ are shown in Table 3 for several $p$-states and their corresponding sub-states. For a given $p$ value, the lowest energy state in Table 3 agrees very well with the value obtained if the current given by Eq.(44) were used, although $V_c(\phi)$ can shift the ordering of states in a way to reorder the ground state moment as seen for the $p=3$ states.
In isolation, this is relatively unimportant. However, in a broader context where the natural temperature scale of a $1000 \AA$ helix is a few $\mu K$, thermodynamic averages of the type
\begin{equation}
\langle T_M \rangle = \sum T_M(E_n) {\rm{exp}}[-E_n/ \tau]
\end{equation}
will necessitate  accounting for proper ordering.

The modifications to the TMs for the upright ($b > a$) coil situation are generally minimal with exceptions only for $p=2$. The flattened coil ($a > b$) results in Table 3 show  a much stronger variation in TM values, consistent with the much larger strength of $V_c(\phi)$ for $a > b$ relative to the converse. A sense of the dependence of TMs on $\omega$ can be gleaned from Table 4 where now $\omega = 8$. Increased variation is seen for both eccentricities, but the flattened coil case demonstrates appreciable deviation from the classical expression.

\section{Conclusions}

In this work a prescription to include curvature at the nanoscale
for particles constrained to toroidal helices was presented, which the authors applied toward a quantum mechanical calculation of toroidal moments.
It is worth emphasizing that the curvature inclusive reduced
dimensionality Schr\"{o}dinger equation developed here
 is driven by an interest in having more tractable,
 effective models for nanomaterials, and is done with the aim of eventually
  confronting experimental data rather than as a purely theoretical exercise.
In that context, the choice to consider helices was driven by their capability of producing
 toroidal moments, which  are currently of both theoretical and practical interest.
 The curvature potential for the helix was derived and shown to be the dominant part of the Hamiltonian for
 lower energy eigenstates of eccentric helices. An intriguing result that arose  here was a
 demonstrated $ab$ asymmetry in  several states of the quantum mechanically calculated  $\textbf{T}^{p\alpha}_M$,
an asymmetry not exhibited in the classical expression of Eq.(43).

The array of  results given in this work was limited to relatively small values of $\omega$ and to
less severe eccentricity because of numerical limitations on
evaluating integrals of the type shown in Eq.(36). The extension to
larger values of $\omega$ were
considered (at least currently) outside the scope of what the
authors were attempting to accomplish. However, preliminary work
gives some indication that \textit{Mathematica} is capable of
performing the necessary integrals, albeit with increased time
expense. It would be of interest to investigate more extreme cases of eccentricity
and loop number given the enhancement of moments already evidenced by larger $\omega$.

Tailoring the response of toroidal helices to electromagnetic
radiation by fabricating objects with curvature as a free
parameter is still well outside the reach of current fabrication
methods. However, the formalism and basis set established here may
serve as means for further investigation of the $\textbf{T}_M \cdot
{\partial \textbf{E} / \partial t}$ interactions relevant to the coupling of toroidal
moments to electromagnetic fields. The extension of the methods here to cases where external vector potentials
are present may be naturally developed from work already done for
tori immersed in arbitrary magnetic fields \cite{encarbB}
and is ongoing with an aim to understanding persistent current effects.

Finally, debate as to whether curvature effects are relevant to, and how they are manifested in,
topologically novel nanostructures may eventually be settled by examining
systems akin to  toroidal helices.
By opting to either include or exclude curvature potentials in  modeling routines,
it may prove true that sensitive quantities like toroidal moments will  provide a
clear signature as to the influence of $V_c$. Work such as that done in this paper may
hopefully contribute to a resolution to the question of how to
properly incorporate twists and turns in the quantum mechanical
description of bent nanostructures.


\renewcommand{\theequation}{A-\arabic{equation}}
\setcounter{equation}{0}  
\section*{Appendix}  

This appendix presents a more complete set of formulae for the circular TH
as well as a corresponding set for the elliptical case.

\subsection*{A.1 The circular case}

\begin{equation}
\hat{\boldsymbol{\theta}}=-{\rm sin}(\omega\phi)\hat{{\boldsymbol {\rho}}}+{\rm cos}(\omega\phi)\hat{{\bf{k}}}
\end{equation}
\begin{equation}
\hat{\textbf{n}}={\rm cos}(\omega\phi)\hat{{\boldsymbol {\rho}}}+{\rm sin}(\omega\phi)\hat{{\bf{k}}}
\end{equation}
\begin{equation}
f(\phi)=(a^2\omega^2+W(\phi)^2)^{1/2}
\end{equation}
\begin{equation}
\hat{\textbf{e}}_2={{W(\phi) \hat{\boldsymbol{\theta}}-a\omega \hat{\boldsymbol{\phi}}}\over{f(\phi)}}
\end{equation}
\begin{equation}
P_1(\phi)=-{{a\omega^2 + W(\phi){\rm cos}(\omega\phi)}\over{f(\phi)^2}}
\end{equation}
\begin{equation}
P_2(\phi)=\frac{{\rm sin}(\omega\phi)}{f(\phi)}\bigg[1+\bigg({a\omega \over{f(\phi)}}\bigg)^2\bigg]
\end{equation}
\begin{equation}
\kappa(\phi)=(P_1^2(\phi)+P_2^2(\phi))^{1/2}
\end{equation}
\begin{equation}
\hat{{\bf{T}}}={{a\omega \hat{\boldsymbol{\theta}}+W(\phi)\hat{\boldsymbol{\phi}}}\over{f(\phi)}}
\end{equation}
\begin{equation}
\hat{{\bf{N}}}={1\over{\kappa(\phi)}}(P_2(\phi) \hat{\textbf{e}}_2+P_1(\phi) \hat{\textbf{n}})
\end{equation}
\begin{equation}
\hat{{\bf{B}}}={1\over{\kappa(\phi)}}(-P_1(\phi) \hat{\textbf{e}}_2+P_2(\phi) \hat{\textbf{n}})
\end{equation}

\subsection*{A.2 The elliptic case}

With $W(\phi)=R+a~{\rm cos}(\omega\phi)$, the equation of the elliptic toroidal helix is $${\bf{r}}(\phi)=W(\phi)\hat{{\boldsymbol {\rho}}}+b~{\rm sin}(\omega\phi)\hat{{\bf{k}}}.$$
The extension of the results of Sec. A.1 to the elliptic case is straightforward:
\begin{equation}
P(\phi)=[a^2~{\rm sin}^2(\omega\phi)+b^2~{\rm cos}^2(\omega\phi)]^{1/2}
\end{equation}
\begin{equation}
\hat{\boldsymbol{\theta}}_E={1 \over{P(\phi)}}[-a~{\rm sin}(\omega\phi)\hat{{\boldsymbol {\rho}}}+b~{\rm cos}(\omega\phi) \hat{{\bf{k}}}]
\end{equation}
\begin{equation}
\hat{\textbf{n}}_E={1 \over{P(\phi)}}[b~{\rm cos}(\omega\phi)\hat{{\boldsymbol {\rho}}}+a~{\rm sin}(\omega\phi) \hat{{\bf{k}}}]
\end{equation}
\begin{equation}
\hat{\textbf{e}}_2={{W(\phi)\hat{\boldsymbol{\theta}}_E+P(\phi)\omega \hat{\boldsymbol{\phi}}}\over{f(\phi)}}
\end{equation}
\begin{equation}
f(\phi)=(P(\phi)^2{\omega}^2+W(\phi)^2)^{1/2}
\end{equation}
\begin{equation}
P_1(\phi)=-{b\over{P(\phi)}}{{a\omega^2 + W(\phi){\rm cos}(\omega\phi)}\over{f(\phi)^2}}
\end{equation}
\begin{equation}
P_2(\phi)=\frac{{\rm sin}(\omega\phi)}{f(\phi)}\bigg({a\over{P(\phi)}}+{{\omega}^2W(\phi)(a^2-b^2){\rm cos}(\omega\phi)+P(\phi)^2a{\omega}^2 \over{f(\phi)^2P(\phi)}}\bigg)
\end{equation}
\begin{equation}
\kappa(\phi)=(P_1(\phi)^2+P_2(\phi)^2)^{1/2}
\end{equation}
\begin{equation}
\hat{{\bf{T}}}={{P(\phi)\omega~\hat{\boldsymbol{\theta}}_E+W(\phi)\hat{\boldsymbol{\phi}}}\over{f(\phi)}}
\end{equation}
\begin{equation}
\hat{{\bf{N}}}={1\over{\kappa(\phi)}}(P_2(\phi) \hat{\textbf{e}}_2+P_1(\phi) \hat{\textbf{n}}_E)
\end{equation}
\begin{equation}
\hat{{\bf{B}}}={1\over{\kappa(\phi)}}(-P_1(\phi) \hat{\textbf{e}}_2+P_2(\phi) \hat{\textbf{n}}_E)
\end{equation}
\newpage
\bibliography{refs2comp}

\begin{thebibliography}{10}
\expandafter\ifx\csname url\endcsname\relax
  \def\url#1{\texttt{#1}}\fi
\expandafter\ifx\csname urlprefix\endcsname\relax\def\urlprefix{URL }\fi
\expandafter\ifx\csname href\endcsname\relax
  \def\href#1#2{#2} \def\path#1{#1}\fi

\bibitem{wegener}
M.~Wegener, S.~Linden, Physics Today 63 (2010) 32.

\bibitem{zhang}
H.~Zhang, S.~W. Chung, C.~A. Mirkin, Nano. Lett. 3 (2003) 43.

\bibitem{shea}
H.R.Shea, R.~Martel, P.~Avouris, Phys. Rev. Lett. 84 (2000) 4441.

\bibitem{lorke}
A.~Lorke, R.~J. Luyken, A.~O. Govorov, J.~P. Kotthaus, Phys. Rev. Lett. 84
  (2000) 2223.

\bibitem{chapblick}
A.~Chaplik, R.~H. Blick, New J. Phys. 6 (2004) 33.

\bibitem{dacosta1}
R.~C.~T. da~Costa, Phys. Rev. A 23 (1981) 1982.

\bibitem{dacosta2}
R.~C.~T. da~Costa, Phys. Rev. A 25 (1982) 2893.

\bibitem{duclosexner}
P.~Duclos, P.~Exner, Rev. Math. Phys. 7 (1995) 73.

\bibitem{ee1}
M.~Encinosa, B.~Etemadi, PRA 58 (1998) 77.

\bibitem{ee2}
M.~Encinosa, B.~Etemadi, Physica B 266 (1998) 361.

\bibitem{jenskoppe}
B.~Jensen, H.~Koppe, Ann. of Phys. 63 (1971) 586.

\bibitem{matsutani1}
S.~Matusani, J. Phys. Soc. Jap. 61 (1991) 55.

\bibitem{matsutani2}
S.~Matsutani, Rev. Math. Phys. 11 (1999) 171.

\bibitem{taira}
H.~Taira, H.~Shima, Surf. Sci. 601 (2007) 5270.

\bibitem{kaelscience}
T.~Kaelberer, V.~Fedetov, N.~Papasimakis, D.~Tsai, N.~Zheludev, Science 330
  (2010) 1510.

\bibitem{afanasiev}
G.~F. Afanasiev, V.~M. Dubovik, G.~Goldoni, F.~Troiani, E.~Molinari, Phys.
  Part. Nucl. 29 (1998) 366.

\bibitem{ceulemans}
A.~Ceulemans, L.~Chibotaru, P.~Fowler, Phys. Rev. Lett. 80 (1998) 1861.

\bibitem{dubovik}
V.~M. Dubovik, V.~V. Tugushev, Phys. Rep. 187 (1990) 145.

\bibitem{papas}
N.~Papasimakis, V.~A. Fedotov, K.~Marinov, N.~I. Zheludev, Phys. Rev. Lett. 103
  (2009) 093901.

\bibitem{naumov}
I.~Naumov, L.~Bellaiche, H.~Fu, Nature 432 (2004) 737.

\bibitem{spaldin}
N.~A. Spaldin, M.~Fiebig, M.~Mostovoy, J. Phys.: Condens. Matter 20 (2008) 1.

\bibitem{sawada}
K.~Sawada, N.~Nagaosa, Phys. Rev. Lett. 95 (2005) 237402.

\bibitem{kibis}
O.~Kibis, M.~Portnoi, Tech. Phys. Lett. 33 (2007) 878.

\bibitem{schujaff}
P.~C. Schuster, R.~L. Jaffe, Ann. Phys. 307 (2003) 132.

\bibitem{burgsjens}
M.~Burgess, B.~Jensen, Phys. Rev. A 48 (1993) 1861.

\bibitem{enchardsoft}
M.~Encinosa, L.Mott, B.~Etemadi, Phys. Scr. 72 (2005) 13.

\bibitem{graustein}
W.~C. Graustein, Differential Geometry, 2nd Edition, Dover, New York, 1962.

\bibitem{arfkenweber}
G.~Arfken, H.~Weber, Mathematical Methods for Physicists, 6th Edition, Elsevier
  Academic Press, Burlington MA, 2002.

\bibitem{encarbB}
M.~Encinosa, Physica E 28 (2005) 209.

\bibitem{encdipole}
M.~Encinosa, J. Comput. Aided Mater. Des. 14 (2006) 65.

\bibitem{grossoparra}
G.~{Grosso}, G.~P. {Parravicini}, Solid State Physics, 1st Edition, Academic
  Press, SanDiego, 2002.

\bibitem{marinov}
K.~Marinov, A.~D. Boardman, V.~A. Fedotov, N.~Zheludev2, N. J. Phys. 9 (2007)
  324.

\end{thebibliography}

\pagebreak
\begin{figure}[h]
\includegraphics[scale=0.74]{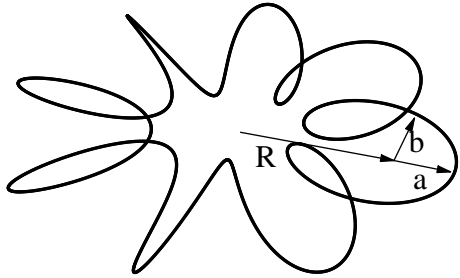}
\begin{center}
\end{center}
\caption{A toroidal helix where $R$ is the distance from the center of the TH to a center of a loop of the TH. The parameter $a$ is the TH's maximum perpendicular horizontal distance from a concentric cylinder of radius $R$. The parameter $b$ is the TH's maximum vertical distance from the x-y plane.}
\end{figure}
\pagebreak
\begin{figure}[h]
\includegraphics[scale=1.0]{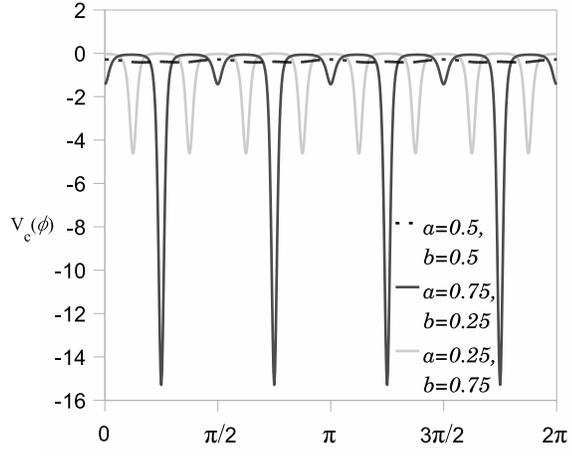}
\begin{center}
\end{center}
\caption{The curvature potential $V_c(\phi)$ in units of $\hbar^2 /(m_e R^2)$ for the case of the circular TH with $R=1$, $\omega=4$, $a=b=0.5$  and two elliptic TH cases: $R=1$, $\omega=4$, $a=0.25$, $b=0.75$ and $R=1$, $\omega=4$, $a=0.25$, $b=0.75$.}
\end{figure}
\pagebreak
\begin{figure}[h]
\begin{center}
\includegraphics[scale=1.00]{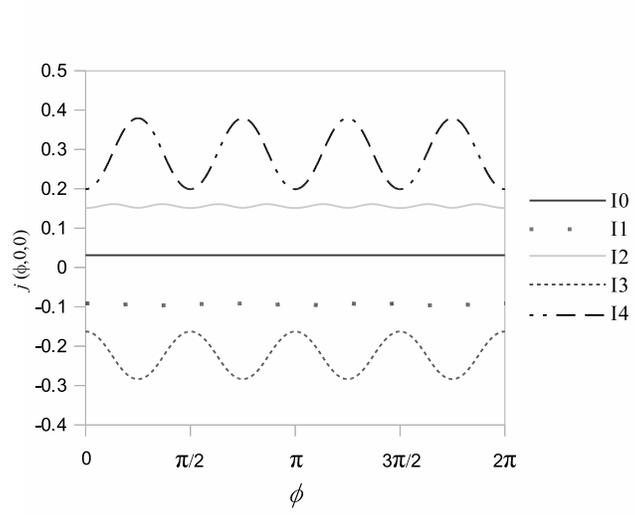}
\end{center}
\caption{${\textit{j~}}^{p\alpha}(\phi,0,0) \equiv I(\phi)$ in units of $q_e \hbar /(m_e R^2)$ for five eigenstates of the circular TH configuration $\omega=4$, $a=0.5$, $b=0.5$, $R=1$, $p=1$ where $V_c(\phi)$ is neglected.}
\end{figure}
\pagebreak
\begin{figure}[h]
\begin{center}
\includegraphics[scale=1.0]{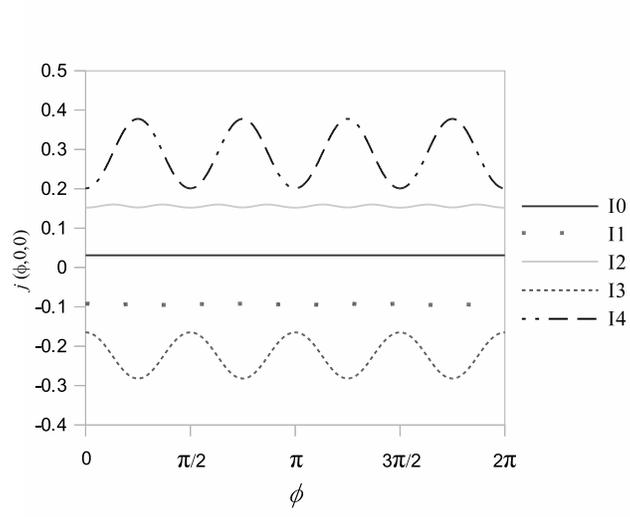}
\end{center}
\caption{${\textit{j}}^{~p\alpha}(\phi,0,0) \equiv I(\phi)$ in units of $q_e \hbar /(m_e R^2)$ for five eigenstates of the circular TH configuration $\omega=4$, $a=0.5$, $b=0.5$, $R=1$, $p=1$ with $V_c(\phi)$. Curvature effects on the current are small in the circular case.}
\end{figure}
\pagebreak
\begin{figure}[h]
\begin{center}
\includegraphics[scale=1.0]{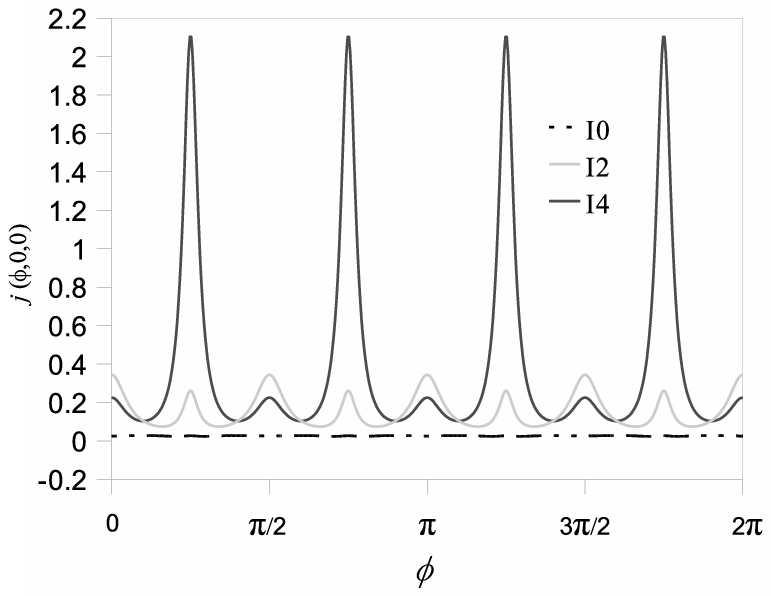}
\end{center}
\caption{${\textit{j}}^{~p\alpha}(\phi,0,0) \equiv I(\phi)$ corresponding to the ground state, and second and fourth excited states in units of $q_e \hbar /(m_e R^2)$, for the elliptic TH configuration $\omega=4$, $a=0.75$, $b=0.25$, $R=1$, $p=1$, without the curvature potential $V_c(\phi)$.}
\end{figure}
\pagebreak
\begin{figure}[h]
\begin{center}
\includegraphics[scale=1.0]{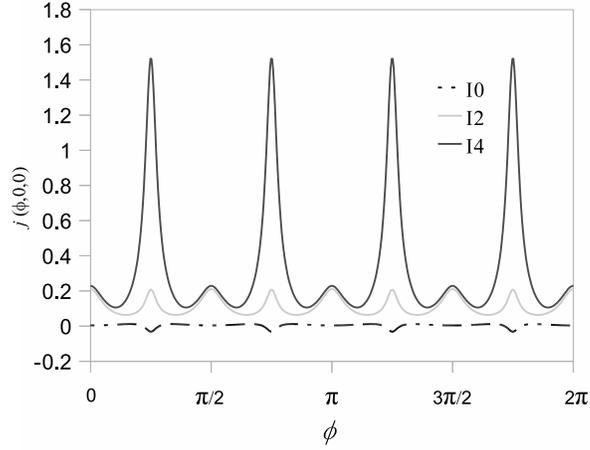}
\end{center}
\caption{${\textit{j}}^{~p\alpha}(\phi,0,0) \equiv I(\phi)$ corresponding to the ground state, and second and fourth excited states in units of $q_e \hbar /(m_e R^2)$, for the elliptic TH configuration $\omega=4$, $a=0.75$, $b=0.25$, $R=1$, $p=1$ inclusive of the curvature potential $V_c(\phi)$. Inclusion of the curvature potential causes a reduction in amplitude for each ${\textit{j}}$.}
\end{figure}
\pagebreak
\begin{figure}[h]
\begin{center}
\includegraphics[scale=1.0]{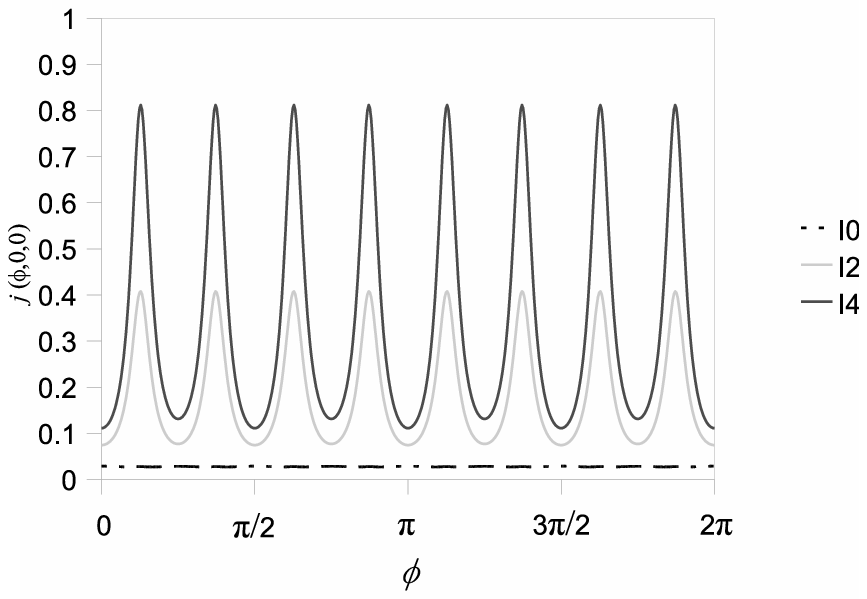}
\end{center}
\caption{${\textit{j}}^{~p\alpha}(\phi,0,0) \equiv I(\phi)$ corresponding to the ground state, and second and fourth excited states in units of $q_e \hbar /(m_e R^2)$ for the elliptic TH configuration $\omega=4$, $a=0.25$, $b=0.75$, $R=1$, $p=1$ without curvature potential.}
\end{figure}
\pagebreak
\begin{figure}[h]
\begin{center}
\includegraphics[scale=1.0]{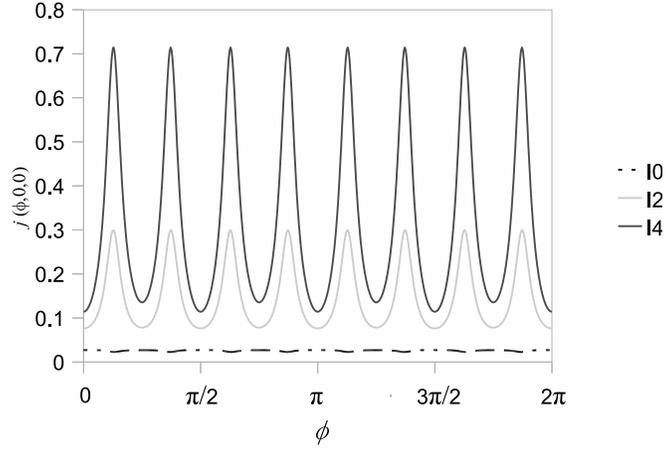}
\end{center}
\caption{${\textit{j}}^{~p\alpha}(\phi,0,0) \equiv I(\phi)$ corresponding to the ground state, and second and fourth excited states in units of $q_e \hbar /(m_e R^2)$ for the configuration $\omega=4$, $a=0.25$, $b=0.75$, $R=1$, $p=1$ inclusive of the curvature potential. When the curvature potential is included in the calculation of $I(\phi)$, the current is reduced in amplitude.}
\end{figure}
\pagebreak
\begin{table}[htbp]
\centering
\begin{tabular}{|c|c|c|c|c|c|}
\hline
& $E_0$ & $E_1$ & $E_2$ & $E_3$ & $E_4$ \\ \hline
& 0.0724 & 1.6369 & 3.0045 & 7.7907 & 10.9186 \\ \hline
$m$ & $C^{(0)}_m$ & $C^{(1)}_m$ & $C^{(2)}_m$ & $C^{(3)}_m$ & $C^{(4)}_m$ \\ \hline
-2 & -0.1055 & -0.1648 & -0.0607 & -0.9761 & 0.0722 \\
-1 & 0.0585 & -0.9762 & -0.1236 & 0.1631 & -0.0428 \\
0 & 0.9822 & 0.0315 & 0.0291 & -0.1020 & 0.1520 \\
1 & 0.0556 & 0.1374 & -0.9702 & 0.0171 & -0.1907 \\
2 & -0.1331 & -0.0087 & -0.1970 & 0.0996 & 0.9662 \\ \hline
\end{tabular}
\caption{Eigenvalues and amplitudes $C^{(\alpha)}_m$ for an $\omega$=6, $a$=0.75, $b$=0.25, $R$=1, $p$=1 elliptic TH neglecting curvature effects.}
\label{}
\end{table}
\begin{table}[htbp]
\centering
\begin{tabular}{|c|c|c|c|c|c|}
\hline
& $E_0$ & $E_1$ & $E_2$ & $E_3$ & $E_4$ \\ \hline
& -1.4739 & -0.0442 & 2.1393 & 5.7258 & 9.2713 \\ \hline
$m$ & $C^{(0)}_m$ & $C^{(1)}_m$ & $C^{(2)}_m$ & $C^{(3)}_m$ & $C^{(4)}_m$ \\ \hline
-2 & 0.0606 & -0.0310 & 0.1231 & -0.9410 & 0.3077 \\
-1 & -0.3794 & -0.7975 & 0.4620 & 0.0389 & -0.0713 \\
0 & 0.8927 & -0.4441 & -0.0424 & 0.0579 & -0.0266 \\
1 & -0.2353 & -0.4071 & -0.8712 & -0.0772 & 0.1179 \\
2 & -0.0062 & 0.0118 & -0.1027 & -0.3220 & -0.9411 \\ \hline
\end{tabular}
\caption{Eigenvalues and amplitudes $C^{(\alpha)}_m$ for an $\omega$=6, $a$=0.75, $b$=0.25, $R$=1, $p$=1 elliptic TH including curvature effects.}
\label{}
\end{table}
\pagebreak
\begin{table}[htbp]
\caption{Toroidal moments for two configurations with $\omega$=4. TM's with and without curvature effects, and classical calculation for each case.}
\begin{minipage}[b]{0.5\linewidth}
\scalebox{0.6}{
\begin{tabular}{|c|c|c|c|c|}
\hline
$\omega$ & $a$ & $b$ &  &  \\ \hline
4 & 0.25 & 0.75 &  &  \\ \hline
p & TM & TM w/$V_c(\phi)$ & ratio & Classical TM \\ \hline
1 & -0.0334 & -0.0317 & 1.0545 & -0.0332 \\
1 & 0.0895 & 0.0718 & 1.2457 & \\
1 & -0.1478 & -0.1308 & 1.1293 & \\
1 & 0.1901 & 0.1837 & 1.0347 &  \\
1 & -0.2401 & -0.2347 & 1.0230 &  \\ \hline
2 & -0.0669 & -0.0551 & 1.2129 & -0.0664 \\
2 & 0.0600 & 0.0524 & 1.1452 & \\
2 & -0.0088 & 0.0193 & -0.4562 &  \\
2 & -0.0031 & -0.0344 & 0.0896 &  \\
2 & -0.2646 & -0.2655 & 0.9967 &  \\ \hline
3 & 0.0288 & 0.0305 & 0.9430 & -0.0995 \\
3 & -0.0993 & -0.0798 & 1.2441 &  \\
3 & 0.1380 & 0.1203 & 1.1479 &  \\
3 & -0.2038 & -0.2052 & 0.9930 &  \\
3 & -0.2888 & -0.2908 & 0.9931 &  \\ \hline
\end{tabular}}
\label{}
\end{minipage}
\begin{minipage}[b]{0.5\linewidth}
\scalebox{0.6}{
\begin{tabular}{|c|c|c|c|c|}
\hline
$\omega$ & $a$ & $b$ &  &  \\ \hline
4 & 0.75 & 0.25 &  &  \\ \hline
p & TM & TM w/$V_c(\phi)$ & ratio & Classical TM \\ \hline
1 & -0.0317 & -0.0068 & 4.6359 & -0.0319 \\
1 & 0.1625 & 0.0627 & 2.5936 &  \\
1 & -0.2697 & -0.1743 & 1.5468 &  \\
1 & 0.1505 & 0.1470 & 1.0240 &  \\
1 & -0.1694 & -0.1863 & 0.9096 &  \\ \hline
2 & -0.0561 & 0.0000 & - & -0.0638 \\
2 & 0.1067 & 0.0157 & 6.7927 &  \\
2 & -0.3206 & -0.1161 & 2.7623 &  \\
2 & 0.1298 & -0.0081 & -15.9737 &  \\
2 & -0.1755 & -0.2072 & 0.8472 &  \\ \hline
3 & 0.0664 & 0.0063 & 10.5665 & -0.0957 \\
3 & -0.0933 & -0.0346 & 2.6957 &  \\
3 & 0.1211 & 0.1092 & 1.1090 &  \\
3 & -0.3893 & -0.3413 & 1.1407 &  \\
3 & -0.1784 & -0.2131 & 0.8372 &  \\ \hline
\end{tabular}}
\label{}
\end{minipage}
\end{table}
\begin{table}[htbp]
\caption{Toroidal moments for two configurations with $\omega$=8. TM's with and without curvature effects, and classical calculation for each case.}
\begin{minipage}[b]{0.5\linewidth}
\scalebox{0.6}{
\begin{tabular}{|c|c|c|c|c|}
\hline
$\omega$ & $a$ & $b$ &  &  \\ \hline
8 & 0.25 & 0.75 &  &  \\ \hline
p & TM & TM w/$V_c(\phi)$ & ratio & Classical TM \\ \hline
1 & -0.0190 & -0.0166 & 1.1485 & -0.0195 \\
1 & 0.1218 & 0.0441 & 2.7619 & \\
1 & -0.1600 & -0.0822 & 1.9477 & \\
1 & 0.2787 & 0.1938 & 1.4383 &  \\
1 & -0.3175 & -0.2351 & 1.3501 &  \\ \hline
2 & -0.0386 & -0.0332 & 1.1615 & -0.0390 \\
2 & 0.1123 & 0.0664 & 1.6917 &  \\
2 & -0.1890 & -0.1430 & 1.3217 &  \\
2 & 0.2635 & 0.2293 & 1.1490 &  \\
2 & -0.3401 & -0.3113 & 1.0923 &  \\ \hline
3 & -0.0582 & -0.0490 & 1.1888 & -0.0584 \\
3 & 0.0952 & 0.0699 & 1.3622 &  \\
3 & -0.2104 & -0.1853 & 1.1354 &  \\
3 & 0.2454 & 0.2238 & 1.0966 &  \\
3 & -0.3598 & -0.3472 & 1.0363 &  \\ \hline
\end{tabular}}
\label{}
\end{minipage}
\begin{minipage}[b]{0.5\linewidth}
\scalebox{0.6}{
\begin{tabular}{|c|c|c|c|c|}
\hline
$\omega$ & $a$ & $b$ &  &  \\ \hline
8 & 0.75 & 0.25 &  &  \\ \hline
p & TM & TM w/$V_c(\phi)$ & ratio & Classical TM \\ \hline
1 & -0.0189 & -0.0096 & 1.9789 & -0.0192 \\
1 & 0.2454 & 0.0560 & 4.3789 &  \\
1 & -0.3224 & -0.1299 & 2.4826 &  \\
1 & 0.3520 & 0.2586 & 1.3612 &  \\
1 & -0.3971 & -0.3162 & 1.2558 &  \\ \hline
2 & -0.0378 & -0.0157 & 2.4149 & -0.0383 \\
2 & 0.2267 & 0.0757 & 2.9954 &  \\
2 & -0.3806 & -0.2231 & 1.7056 &  \\
2 & 0.3382 & 0.2988 & 1.1315 &  \\
2 & -0.4284 & -0.4176 & 1.0257 &  \\ \hline
3 & -0.0565 & -0.0138 & 4.1053 & -0.0575 \\
3 & 0.1951 & 0.0614 & 3.1800 &  \\
3 & -0.4221 & -0.2692 & 1.5680 &  \\
3 & 0.3116 & 0.2627 & 1.1858 &  \\
3 & -0.4509 & -0.4640 & 0.9718 &  \\ \hline
\end{tabular}}
\label{}
\end{minipage}
\end{table}
\end{document}